\documentclass{article}
\usepackage{amssymb}
\usepackage{amsfonts}
\usepackage[a4paper]{geometry}
\usepackage{amsmath}

\begin{document}

\title{ Quantum Dynamics of a Particle in a Linear Potential: Invariant
Operator Approach and Discrete Spectrum Solutions}
\author{Mustapha Maamache$^{1\dagger \thanks{%
email: maamache\_m@yahoo.fr }}$ and Aymen Bendjoudi$^{1\text{,}2}$ \\
$^{1}$Laboratoire de Physique Quantique et Syst\`{e}mes Dynamiques,\\
Facult\'{e} des Sciences, Universit\'{e} Ferhat Abbas S\'{e}tif 1, \\
S\'{e}tif 19000, Algeria\\
$^{2}$Department of Physics, \\
Graduate School of Natural and Applied Sciences,\\
Eski\c{s}ehir Technical University, 26555, Eski\c{s}ehir, Turkey.\ \ \ }
\maketitle

\begin{abstract}
We investigate the quantum dynamics of a particle subjected to a linear
potential using the Lewis--Riesenfeld invariant operator method. Starting
from the time-dependent Schr\"{o}dinger equation associated with a constant
external force, we construct the most general Hermitian quadratic invariant
and derive the corresponding coupled differential equations for its
time-dependent coefficients. By means of an appropriate sequence of unitary
transformations, the invariant operator is reduced to the form of a harmonic
oscillator Hamiltonian. This reduction enables a clear classification of the
system according to the sign of the conserved quantity $\omega ^{2}$.
Particular attention is devoted to the physically relevant case $\omega ^{2}$
$>0$, which yields a discrete eigenspectrum. Explicit analytical expressions
for the invariant coefficients, the displacement parameters, and the
transformed wave functions are obtained. The resulting formalism provides an
exact quantum description of a particle under a constant force and
establishes a direct connection between invariant theory and harmonic
oscillator quantization.

PACS: 03.65.Ca, 03.65.Ta , 03.65.Vf, 03.65.Nk

In memory of Maamache Leulmi-Amar and Djabou Zoulikha, beloved father and
mother of Mustapha Maamache.
\end{abstract}

\section{Introduction}

Quantum mechanics provides the fundamental framework for describing the
behavior of microscopic particles under the influence of external forces and
potentials. Among the various quantum systems studied in theoretical
physics, the motion of a particle subjected to a linear potential occupies a
central position because of its rich mathematical structure and broad range
of physical applications.

Physically, a linear potential represents the quantum analogue of a particle
moving under a constant external force, such as a charged particle placed in
a uniform electric field. Owing to its fundamental nature, this model
appears in several branches of modern physics, including condensed matter
physics, quantum information theory, and astrophysics, where the confinement
and manipulation of particles by external fields play an essential role.

The dynamics of the system are governed by the time-dependent Schr\"{o}%
dinger equation

\begin{equation}
i\hbar \frac{\partial }{\partial t}\psi (q,t)=H\text{ }\psi (q,t)  \label{1}
\end{equation}%
{} with the Hamiltonian

\begin{equation}
H\left( q,p\right) =\frac{1}{2m}p^{2}-Fq,  \label{2}
\end{equation}%
where $m$ denotes the particle mass and $F$ is the constant external force.

Unlike the harmonic oscillator, the linear potential does not possess a
discrete spectrum that can be expressed through elementary functions.
Instead, the stationary solutions are written in terms of Airy functions 
\cite{7},

\begin{equation}
\psi (\xi )=C_{1}\,Ai(\xi )+C_{2}\,Bi(\xi ),\ \xi =\left( 2mF\right)
^{1/2}(q-\frac{E}{F}),  \label{2'}
\end{equation}%
where $Ai(\xi )$ and $Bi(\xi )$ are Airy functions, while $C_{1},C_{2}$ are
constants determined by the boundary conditions. Since the function $Bi(\xi
) $ diverges as $\xi \rightarrow \infty $, it violates the physical
requirement that the wave function remains finite. Consequently, only $%
Ai(\xi )$ corresponds to physically acceptable solutions.

The time evolution of a quantum system subjected to a spatially uniform
time-dependent force has also attracted considerable attention \cite%
{8,9,10,11,12,13,15,16,17}. The exact propagator for this system has long
been known, as well as a class of exact solutions usually referred to as
Volkov solutions. More nvestigations have focused on the derivation of exact
wave-packet solutions and on the analysis of their physical properties.

The general quantum-mechanical properties of dynamical systems can be
investigated using the invariant operator theory introduced by Lewis and
Riesenfeld \cite{18}. In this approach, the eigenfunctions of the invariant
operator coincide with the wave functions of the Schr\"{o}dinger equation up
to time-dependent phase factors. The invariant formalism therefore provides
a powerful framework for obtaining exact quantum solutions.

In the present work, we focus on wave solutions characterized by a discrete
eigenspectrum. In particular, we investigate the case corresponding to
positive frequency, for which the transformed invariant operator reduces to
a harmonic oscillator with discrete eigenvalues.

\section{ Lewis--Riesenfeld Invariant Formalism and Discrete Spectrum
Structure}

According to the Lewis--Riesenfeld theory, a Hermitian operator $I(t)$ is
called an invariant if it satisfies the Liouville--von Neumann equation 
\begin{equation}
\frac{dI}{dt}=\frac{\partial I}{\partial t}+\frac{1}{i\hbar }\left[ I,H%
\right] =0,  \label{3}
\end{equation}%
if $\varphi _{\lambda }(q,t)$ is an eigenfunction of $I(t)$ with a
time-independent eigenvalue $\zeta _{\lambda }$,%
\begin{equation}
I(t)\varphi _{\lambda }(q,t)=\zeta _{\lambda }\text{ }\varphi _{\lambda
}(q,t),  \label{3"}
\end{equation}%
(the subscript $\lambda $ depends on the nature of the spectrum (discrete
for $\lambda $ $=n$) or (continuous for $\lambda =k$) then a solution of the
Schr\"{o}dinger equation can be written as 
\begin{equation}
\psi _{\lambda }(q,t)=\exp \left[ i\alpha _{\lambda }\left( t\right) \right]
\varphi _{\lambda }(q,t)  \label{3'}
\end{equation}%
where the phase $\alpha _{\lambda }\left( t\right) $ satisfies

\begin{equation}
\hbar \dot{\alpha}_{\lambda }\left( t\right) \varphi _{\lambda }=\left[
i\hbar \frac{\partial }{\partial t}-H\right] \varphi _{\lambda }.  \label{4}
\end{equation}

\bigskip\ To investigate the quantum dynamics of a particle subjected to the
linear potential, we consider the most general quadratic Hermitian invariant 
\begin{equation}
I\left( t\right) =\ \ \left[ A\left( t\right) p^{2}+B\left( t\right) \left(
pq+qp\right) +C\left( t\right) q^{2}+g\left( t\right) q+K\left( t\right)
p+\gamma \left( t\right) \right] ,  \label{e}
\end{equation}%
where the coefficients\ \ $A\left( t\right) $,\ $B\left( t\right) $, $C(t),$ 
$g\left( t\right) $, $K\left( t\right) $and $\gamma \left( t\right) $\ are
real functions of time.\ \ \ \ \ \ \ \ \ \ \ \ \ \ \ \ \ \ \ \ \ \ \ \ \ \ \
\ \ \ \ \ \ \ \ \ \ \ \ \ \ \ \ \ \ \ \ \ \ \ \ \ \ \ \ \ \ \ \ \ \ \ \ \ \
\ \ \ \ \ \ \ \ \ \ \ \ \ \ \ \ \ 

Substituting Eq. (\ref{e}) into Eq. (\ref{3}) yields the coupled
differential system

\begin{equation}
\left\{ 
\begin{array}{c}
\overset{\cdot }{A}\left( t\right) +\frac{2B\left( t\right) }{m}=0 \\ 
\ \overset{\cdot }{B}\left( t\right) +\frac{C\left( t_{0}\right) }{m}=0 \\ 
\overset{\cdot }{C}\left( t\right) =0 \\ 
\ \overset{\cdot }{g}\left( t\right) +2fB\left( t\right) =0 \\ 
\overset{\cdot }{K}\left( t\right) +\left( 2fA\left( t\right) +\frac{g\left(
t\right) }{m}\right) =0 \\ 
\overset{\cdot }{\gamma }\left( t\right) +fK\left( t\right) =0.%
\end{array}%
\right.  \label{5}
\end{equation}%
The third equation implies immediately that $C(t)$ is constant:%
\begin{equation}
C\left( t\right) =C_{0},  \label{M}
\end{equation}%
where $C_{0}$ is a constant. Integrating successively, one obtains%
\begin{equation}
\left\{ 
\begin{array}{c}
B\left( t\right) =B_{0}-\frac{C_{0}}{m}(t-t_{0}), \\ 
A\left( t\right) =A_{0}-\frac{2B_{0}}{m}(t-t_{0})+\frac{C_{0}}{m^{2}}%
(t-t_{0})^{2}, \\ 
g\left( t\right) =g_{0}-2fB_{0}(t-t_{0})+\frac{C_{0}f}{m}(t-t_{0})^{2}, \\ 
K\left( t\right) =K_{0}-\left( 2fA_{0}+\frac{g_{0}}{m}\right) (t-t_{0})+%
\frac{3fB_{0}}{m}(t-t_{0})^{2}-\frac{C_{0}f}{m^{2}}(t-t_{0})^{3}, \\ 
\gamma \left( t\right) =\gamma _{0}-fK_{0}(t-t_{0})+f\frac{\left( 2fA_{0}+%
\frac{g_{0}}{m}\right) }{2}(t-t_{0})^{2}-\frac{f^{2}B_{0}}{m}(t-t_{0})^{3}+%
\text{ }\frac{C_{0}f^{2}}{4m^{2}}(t-t_{0})^{4}.%
\end{array}%
\right.  \label{L}
\end{equation}%
An important conserved quantity follows naturally from these relations:

\begin{equation}
A\left( t\right) C\left( t\right) -B^{2}\left( t\right) =\omega ^{2},
\label{P}
\end{equation}%
with%
\begin{equation}
\omega ^{2}=A_{0}C_{0}-B_{0}^{2}.  \label{Q}
\end{equation}%
To simplify the invariant operator, we introduce the time-dependent unitary
transformation $U(t)$ = $U_{1}(t)U_{2}(t)D^{+}\left( \alpha \left( t\right)
\right) $ such that%
\begin{equation}
\varphi _{\lambda }\left( q,t\right) =D\left( \alpha \left( t\right) \right)
U_{2}^{+}(t)U_{1}^{+}(t)\tilde{\varphi}_{\lambda }\left( q\right) ,
\label{B}
\end{equation}%
where 
\begin{equation}
\left\{ 
\begin{array}{c}
U_{1}(t)=e^{\frac{i}{4\hbar }(qp+pq)\ln (2A\left( t\right) )}, \\ 
U_{2}(t)=e^{\frac{i}{2\hbar }\frac{B\left( t\right) }{A\left( t\right) }%
q^{2}}, \\ 
D\left( \alpha \left( t\right) \right) =e^{-\frac{i}{\hbar }(\eta q-\mu p)}.%
\end{array}%
\right.  \label{9}
\end{equation}%
The displacement parameters are given by 
\begin{equation}
\left\{ 
\begin{array}{c}
\mu \left( t\right) =\frac{A\left( t\right) g\left( t\right) -K\left(
t\right) B\left( t\right) }{2\left( A\left( t\right) C\left( t\right)
-B\left( t\right) ^{2}\right) } \\ 
\eta \left( t\right) =\frac{C\left( t\right) K\left( t\right) -g\left(
t\right) B\left( t\right) }{2\left( A\left( t\right) C\left( t\right)
-B\left( t\right) ^{2}\right) }%
\end{array}%
\right.  \label{10}
\end{equation}%
Under these transformations, the coordinate and momentum operators transform
according to 
\begin{equation}
\left\{ 
\begin{array}{c}
\begin{array}{c}
\begin{array}{c}
U_{1}(t)qU_{1}^{+}(t)=\sqrt{2A\left( t\right) }\text{ }q, \\ 
U_{1}(t)pU_{1}^{+}(t)=\frac{1}{\sqrt{2A\left( t\right) }\text{ }}p, \\ 
U_{2}(t)pU_{2}^{+}(t)=p\text{ }-\frac{B\left( t\right) }{A\left( t\right) }q,%
\end{array}
\\ 
D^{+}\left( \alpha \left( t\right) \right) qD\left( \alpha \left( t\right)
\right) =q-\mu \left( t\right) ,%
\end{array}
\\ 
D^{+}\left( \alpha \left( t\right) \right) pD\left( \alpha \left( t\right)
\right) =p-\eta \left( t\right) .%
\end{array}%
\right.  \label{A}
\end{equation}

\bigskip After straightforward calculations, the transformed invariant
operator becomes%
\begin{equation}
I^{\prime }(t)=U(t)I(t)U^{+}(t)=\frac{1}{2}\left( p^{2}+4\omega
^{2}q^{2}\right) +\Lambda  \label{D}
\end{equation}

where 
\begin{equation}
\Lambda =A\left( t\right) \eta \left( t\right) ^{2}+2B\left( t\right) \eta
\left( t\right) \mu \left( t\right) +C\left( t\right) \mu \left( t\right)
^{2}-g\left( t\right) \mu \left( t\right) -K\left( t\right) \eta \left(
t\right) +\gamma \left( t\right) .  \label{42}
\end{equation}%
i.e., $U(t)$ brings any solution of the operator eigenvalue equation (\ref%
{3"}) into a solution of the operator eigenvalue equation%
\begin{equation}
I^{\prime }(t)\tilde{\varphi}_{\lambda }\left( q\right) =\zeta _{\lambda }%
\tilde{\varphi}_{\lambda }\left( q\right) ,  \label{C}
\end{equation}%
The transformed invariant operator is defined by and%
\begin{equation}
\varphi _{\lambda }\left( q,t\right) =D\left( \alpha \left( t\right) \right)
U_{2}^{+}(t)U_{1}^{+}(t)\tilde{\varphi}_{\lambda }\left( q\right) ;
\label{8}
\end{equation}

After straightforward calculations, the transformed invariant operator $%
I^{\prime }(t)=U(t)I(t)U^{+}(t)$ takes the form

\begin{equation}
I^{\prime }=\frac{1}{2}\left( p^{2}+4\omega ^{2}q^{2}\right) +\Lambda
\end{equation}%
Using the explicit expressions of the coefficients, the constant term $%
\Lambda $ reduces to the explicit expressions of the coefficients, Eq. (\ref%
{42}) simplifies to 
\begin{equation}
\Lambda =\frac{-A_{0}g_{0}^{2}-C_{0}K_{0}^{2}+2B_{0}g_{0}K_{0}+4A_{0}C_{0}%
\gamma _{0}-4B_{0}^{2}\gamma _{0}}{4\left( A_{0}C_{0}-B_{0}^{2}\right) }.
\label{E}
\end{equation}%
Without loss of generality, we choose $g_{0}=K_{0}=\gamma _{0}=0$, so that
so that the invariant operator finally reduces to

\begin{equation}
I^{\prime }(t)=\frac{1}{2}\left( p^{2}+4\omega ^{2}q^{2}\right) .  \label{I'}
\end{equation}

The eigenvalue equation associated with Eq. (\ref{I'}) is therefore 
\begin{equation}
\frac{1}{2}\left[ -\hbar ^{2}\frac{\partial ^{2}}{\partial q^{2}}+4\omega
^{2}q^{2}\right] \tilde{\varphi}_{\lambda }\left( q\right) =\zeta _{\lambda }%
\tilde{\varphi}_{\lambda }\left( q\right) ,  \label{os}
\end{equation}%
which is exactly the stationary Schr\"{o}dinger equation of the
one-dimensional harmonic oscillator.

The sign of $\omega ^{2}$ determines the spectral structure of the system:

$\cdot $ $\omega ^{2}$ $>0$: discrete spectrum,

$\cdot $ $\omega ^{2}$ $=0$: continuous spectrum,

$\cdot $ $\omega ^{2}$ $<0$: continuous spectrum.

In the present work, we focus exclusively on the physically relevant regime $%
\omega ^{2}$ $>0$, where the transformed invariant describes an oscillatory
system with quantized eigenvalues. The normalized eigenfunctions are then
given by

\begin{equation}
\tilde{\varphi}_{n}\left( x\right) =\left( \frac{2\omega }{\hbar }\right) ^{%
\frac{1}{4}}\frac{1}{\sqrt{\sqrt{\pi }2^{n}n!}}e^{-\frac{\omega x^{2}}{\hbar 
}}H_{n}\left( \left( \frac{2\omega }{\hbar }\right) ^{\frac{1}{2}}q\right)
,\qquad
\end{equation}%
with eigenvalues%
\begin{equation}
\zeta _{n}=2\hbar \omega \left( n+\frac{1}{2}\right) ,\text{ \ \ \ \ \ }%
n=0,1,2,...
\end{equation}%
where $H_{n}$ are the Hermite polynomials.

The displacement parameters introduced through the unitary transformation $%
D\left( \alpha \left( t\right) \right) $ satisfy

$\ \ \ \ \ \ \ \ \ \ \ \ \ \ \ \ \ \ \ \ \ \ \ \ \ \ \ \ \ \ \ \ \ \ \ \ \ \
\ \ \ \ \ \ \ \ \ \ \ \ \ \ \ \ \ \ \ \ \ \ \ \ \ \ \ \ \ $%
\begin{equation}
\left\{ 
\begin{array}{c}
\mu \left( t\right) =\mu _{0}-\frac{\eta _{0}}{m}(t-t_{0})+\frac{f}{2m}%
(t-t_{0})^{2} \\ 
\eta \left( t\right) =\eta _{0}-f(t-t_{0}).%
\end{array}%
\right.  \label{m}
\end{equation}%
Differentiation gives\qquad 
\begin{equation}
\overset{\cdot }{\mu }\left( t\right) =\frac{\eta \left( t\right) }{m}\text{%
, \ \ \ }\overset{\cdot }{\eta }\left( t\right) =-f\text{\ ,\ \ \ }
\label{mm}
\end{equation}%
which reproduce exactly the classical equations of motion for a particle
subjected to a constant external force. Hence, the classical dynamics emerge
naturally from the invariant operator formalism and the associated unitary
transformations.

\section{ Conclusion}

In this work, we studied the quantum dynamics of a particle subjected to a
linear potential using the Lewis--Riesenfeld invariant operator method.
Starting from the most general quadratic Hermitian invariant, we derived the
complete set of differential equations governing the invariant coefficients
and obtained their exact analytical solutions.

Through a sequence of time-dependent unitary transformations, the invariant
operator was reduced to the Hamiltonian form of a harmonic oscillator. This
transformation revealed the existence of a conserved quantity $\omega ^{2}$ $%
>0$, whose sign determines the spectral nature of the system.

Special attention was devoted to the case $\omega ^{2}$ $>0$, corresponding
to a discrete spectrum. In this regime, the transformed invariant admits
harmonic oscillator eigenfunctions and quantized eigenvalues. Explicit
expressions for the displacement parameters and associated phase structure
were also obtained.

The present formulation provides an exact and elegant framework for studying
quantum systems subjected to constant external forces and establishes a
direct correspondence between linear-potential dynamics and harmonic
oscillator quantization through invariant theory.

\subparagraph{ $\circ $ Author Contribution declaration in the manuscript}

The study was conceived and schemed by Maamache Mustapha.

The mathematical evaluations in the text were performed by Aymen Bendjoudi
and Maamache Mustapha .

The paper was written by Mustapha Maamache and the final correction of the
paper was done by Aymen Bendjoudi

\subparagraph{$\circ $ Conflicts of interests}

.The authors declare no conflict of interest.

$\circ $ \textbf{Funding Declaration}

No funding Declaration

\end{document}